 \documentclass[nohyper,12pt,letterpaper]{JHEP3}
 \usepackage{epsfig}
 
 \def\bR{{\mathbb R}}

 \title{Flavor from M5-branes}
 \author{Bartomeu  Fiol\\
 Departament de F{\'\i}sica Fonamental i \\Institut de Ci{\`e}ncies del Cosmos, 

Universitat de Barcelona,

Mart{\'\i}\ i Franqu{\`e}s 1, 08193 Barcelona, Catalonia, Spain.\\

\email{bfiol@ub.edu}}

\abstract{We study various aspects of the defect conformal field theory that arises when placing a single M5-brane probe in $AdS_4\times S^7$. We derive the full set of fluctuation modes and dimensions of the corresponding dual operators. We argue that the latter does not depend on the presence of a non-trivial magnetic flux on the M5-brane world-volume. Finally we give a mass to the hypermultiplet living on the defect, and compute the resulting mesonic spectrum. }

\begin{document}

\section{Introduction}
An interesting way to construct novel field theories with conformal symmetry is to consider a known CFT in $d$ spacetime dimensions, and introduce a submanifold on which the fields of the CFT must satisfy some boundary conditions \cite{Cardy:1984bb}. The introduction of this defect clearly breaks the original $SO(2,d)$ conformal group, but by suitably choosing the geometry of the boundary, and the boundary conditions imposed on the fields, it is possible to preserve a smaller conformal symmetry. In physical applications, there are often additional degrees of freedom living only on this submanifold (the 'defect«). Such defect theories have been considered in a variety of physical applications in condensed matter.

Defect conformal field theories can be realized within the framework of the AdS/CFT correspondence in a quite straightforward manner, by embedding a suitable probe brane in a type II or M theory solution with an AdS factor \cite{Karch:2000gx, Bachas:2000fr, Skenderis:2002vf, Yamaguchi:2003ay, Lunin:2007ab}. This realization of the dCFT brings in geometrical intuition, and it has proved to be very useful in studying these field theories \cite{Bachas:2001vj, DeWolfe:2001pq, Constable:2002xt}. While in this context the AdS/CFT correspondence has been used mostly to understand better the field theories, there has been also work in the other direction, trying to use these theories to elucidate aspects of gravity, like the possibility of localizing gravity on a brane \cite{Karch:2001cw, Aharony:2003qf}, or discussing possible holographic descriptions of eternal inflation \cite{Fiol:2010wf}.

One can also study non-conformal phases of a defect field theory, either by adding deformations that break conformal invariance, or by simply starting with a brane configuration that is holographically dual to a non-conformal field theory \cite{Arean:2006pk, Myers:2006qr, Arean:2006vg, Arean:2007nh}. Furthermore, by considering non-supersymmetric probe branes, one can study the phase diagram of these theories \cite{Hung:2009qk}. 

While most of the work on this topic has dealt with defect field theories coming from D-branes in type II backgrounds, there has also been some work on defect CFTs coming from M-branes in 11d SUGRA backgrounds. Relevant probe branes were presented in \cite{Yamaguchi:2003ay, Lunin:2007ab}, and partial results for the resulting defect CFTs and its deformations away from conformal invariance have appeared in \cite{Arean:2006pk, Myers:2006qr, Arean:2007nh, Ammon:2009wc}. However, as far as I am aware, there is no available example of a fully worked-out analysis of defect CFTs realized by M-branes: for instance, in \cite{Arean:2006pk, Myers:2006qr, Arean:2007nh} the fluctuation analysis was restricted to a subset of M5 world-volume modes, while \cite{Ammon:2009wc} aimed at providing a Lagrangian description of various defects for ABJM \cite{Aharony:2008ug} CFTs , but encountered difficulties precisely when these defects are realized by M5-branes. The aim of this note is to improve this state of affairs, by working out a particular example, and laying the groundwork for future extensions. Specifically we consider the defect CFT obtained by placing a single M5 probe brane in $AdS_4\times S^7$. The defect theory is a 1+1 ${\cal N}=(4,4)$ non-chiral CFT with a 1+1 ${\cal N}=(4,4)$ hypermultiplet. We obtain the full spectrum of dimensions of operators in short multiplets. We then add mass to the hypermultiplet,  and compute the resulting spectrum of mesons. We leave the study of the phase diagram of this theory for future work.

The present paper is organized as follows. In section two we introduce the probe brane that is holographically dual to the defect CFT, namely an M5-brane in $AdS_4\times S^7$. We consider the possibility of a non-trivial magnetic flux on the brane, since in our recent work \cite{Fiol:2010wf} applying this system to study the holographic multiverse proposal, it was crucial to turn on such flux. In section three we expand the Lagrangian around this solution, and write down the fluctuation Lagrangian. We then argue that the spectrum of fluctuations depends trivially on the presence of magnetic flux, with the interesting consequence that the dimensions of the dual CFT operators are independent of it. We go on to actually compute the spectrum of such fluctuations; this computation shares many generic features with similar ones performed to determine the fluctuation spectrum in supergravity and in various probe brane setups. From the $AdS_3$ masses of the fluctuations it is then immediate to derive the dimensions of the dual operators, using the standard AdS/CFT dictionary. Finally, in section four we add a mass to the hypermultiplet living on the hypermultiplet; this breaks conformal invariance and the resulting spectrum consists now on a discrete set of mesons. There were partial results in the literature for this case  \cite{Arean:2006pk, Myers:2006qr}, but those papers studied only the mesons coming from fluctuations of the directions transverse to $S^3$ in $S^7$. We complete these works by computing the full set of mesons, coming from the fluctuations of all world-volume fields.

\section {Preliminaries}
In this section we introduce the probe brane solution whose fluctuations we will be discussing in the rest of the paper. We start by writing the supergravity background in the most convenient way for our purposes, and then we display the probe brane solution. This solution has already appeared in the literature, so we will be brief, since our purpose is mostly to set the notation.

\subsection{The background}
The background we will be considering is the $AdS_4\times S^7$ solution of 11D SUGRA, where $R_{AdS_4}=2R_{S^7}$.  We write the $AdS_4\times S^7$ solution in a way that makes manifest a 2+1 Minkowski boundary,
$$
ds^2=\frac{r^4}{R^4}dx_{1,2}^2+\frac{R^2}{r^2}d\vec r^2 
$$
$$
 C_{(3)}=\frac{r^6}{R^6}dx^0\wedge dx^1\wedge dx^2
$$
in the previous formulae, $\vec r$ is an 8-dimensional vector in $\bR^8$. In order to present the particular M5-brane embedding we will be discussing, it is convenient to split this eight dimensional vector into two four dimensional ones, $\vec y$ and $\vec z$, with $\rho$ the norm of $\vec y$. The background metric reads then
\begin{equation}
ds^2=\frac{(\rho^2+\vec z^2)^2}{R^4}dx^2_{1,2}+\frac{R^2}{\rho^2+\vec z^2}\left(d\rho^2+\rho^2 d\Omega_3^2+d\vec z^2\right)
\label{backmet}
\end{equation}
The reason to perform this split is to make manifest a choice of an $S^3$ inside $S^7$ that will become handy when we next discuss the probe embedding.

\subsection{The probe}
The probe we consider is an M5-brane. The world-volume content of a single M5 brane is a 6d (2,0) tensor multiplet, which includes a 2-form $B$ whose 3-form field-strength $F=dB$ is self-dual. As it is well known, writing an action for a self-dual field can be quite delicate. In the case at hand, a way to overcome these difficulties involves adding an auxiliary scalar field $a(\xi)$, with a non-polynomial action. We follow this route and as the world-volume action of the M5 brane we will take the PST action \cite{Pasti:1997gx},
$$
S_{PST}=T_{M5}\int d^6\xi \left[-\sqrt{-|G_{ij}+\check H_{ij}|}+\frac{\sqrt{-|G|}}{4\partial a \cdot \partial a}
\partial _i a (*H)^{ijk}H_{jkl}\partial^l a\right]
 $$
\begin{equation}
+T_{M5}\int \frac{1}{2}F\wedge P[C^{(3)}]+P[C^{(6)}]
\label{pstact}
\end{equation}
where $G$ is the world-volume metric, $H$ is a 3-form that modifies $F$ with the pull-back of the background form $C^{(3)}$
$$
H=F-P[C^{(3)}]
$$
and finally
\begin{equation}
\check H^{ij}=\frac{1}{3!\sqrt{-|G|}}\frac{1}{\sqrt{-(\partial a)^2}}\epsilon^{ijklmn}\partial _k a H_{lmn}
\label{hhat}
\end{equation}
Our index conventions are as follows: $i,j=0,1,\dots,5$ denote world-volume coordinates. As we will describe below, we will fix the gauge for the auxiliary field $a(\xi)$ by setting $a(\xi)=\xi^1$. This singles out this direction and it is convenient to introduce indices $a,b=0,2,3,4,5$ that run over the rest of world-volume coordinates. The solutions that we consider have an induced metric of the form $AdS_3\times S^3$, and we use $\mu,\nu$ to denote $AdS_3$ coordinates and $\alpha,\beta$ to denote $S^3$ coordinates.

Let's now describe the M5 embedding in $AdS_4\times S^7$ that we will be considering. The world-volume coordinates $\xi^i$ are identified with $x^0,x^1, \rho$ and the coordinates of the $S^3$ in $S^7$ singled out in (\ref{backmet}). Then, as ansatz for the other fields we take 
$$
x^2=x^2(\rho),\hspace{1cm}\vec z=0, \hspace{1cm} F=q\; \hbox{vol}(S^3)
$$
We still need to fix the gauge for the world-volume non-dynamical field $a(\xi)$. As is obvious from the way it appears in the action, it is not possible to set it to a constant value. Following \cite{Pasti:1997gx, Arean:2007nh} we fix the gauge by setting $a=x^1$. The price we pay is that we lose manifest  world-volume coordinate covariance; this will be reflected in the fact that in some expressions below the world-volume indices $i,j$ will split as 1 and $a,b$.

The solution found in \cite{Arean:2007nh} for this ansatz is
\begin{equation}
x^2(\rho)=x^2(\infty)+\frac{q}{2\rho^2}
\label{probesol}
\end{equation}
It can be checked \cite{Arean:2007nh} that this solution is 1/2 BPS. In the 2+1 boundary, this probe brane ends at a 1+1 domain wall  placed at $x^2=x^2(\infty)$. The induced metric on the M5-brane world-volume is
\begin{equation}
ds^2_{M5}=G_{ij}d\xi^id\xi^j=\frac{\rho^4}{R^4}dx^2_{1,1}+R^2\left(1+\frac{q^2}{R^6}\right) \frac{d\rho^2}{\rho^2}+R^2 d\Omega_3^2
\label{indmet}
\end{equation}
it is of the form $AdS_3(R_{eff}/2)\times S^3(R)$, with
\begin{equation}
R_{eff}^2=R^2\left(1+\frac{q^2}{R^6}\right)
\label{radieff}
\end{equation}
We close this section by discussing the symmetries preserved by this solution. Before adding the probe brane the susy algebra of the $AdS_4\times S^7$ solution is OSp(8$\mid$4,$\bR$), whose bosonic subalgebra is $SO(3,2)\times SO(8)$. The probe breaks the bosonic symmetry to  $SO(2,2)\times SO(4)\times SO(4)$, and we use the results of  \cite{{D'Hoker:2008ix}}, who classify half-BPS solutions of M-theory to deduce the full supersymmetry algebra preserved by this probe. This probe corresponds to case 5 of table 5 in \cite{{D'Hoker:2008ix}}, and the backreacted solution of the probe brane considered above corresponds to case VII in table 12 of  \cite{D'Hoker:2008ix}, so the superalgebra is $Osp(4|2,\bR) \oplus OSp(4|2,\bR)$, which indeed has bosonic subalgebra $SO(2,2)\times SO(4)\times SO(4)$.

\section{Fluctuation analysis}
In the previous section we have written an $M5$ probe brane embedding in $AdS_4\times S^7$. By the usual $AdS/dCFT$ arguments \cite{Karch:2000gx, DeWolfe:2001pq}, it is expected to be dual to a 2+1 CFT modified by a codimension one defect placed at $x^2=x^2(\infty)$. In this section we are going to study the spectrum of  small fluctuations of the bosonic world-volume fields of the $M5$ brane about this solution, which will enable us to deduce the dimensions of operators in short multiplets of the dual theory.

We expand the world-volume fields around their value at the solution (\ref{probesol}),
$$
x^2=x^2_{sol}+h(\xi)\hspace{.5cm} z^m=0+\chi^m(\xi)\hspace{.5cm}B_{ij}=B_{ij}^{sol}+b_{ij}(\xi)
$$
so $h,\chi^m,b$ are the fluctuating fields. Since $B$ only appears in the action through $F=dB$, its fluctuation will appear only through $f\equiv db$. To derive the fluctuation Lagrangian we plug these expressions into the action (\ref{pstact}) and expand up to quadratic terms. We will now outline the computation, presenting only the key intermediate steps. 

First we consider the term with the square root in the action (\ref{pstact}). We will use the labels $s$ and $f$ ro refer respectively to the solution and fluctuation pieces of various tensors. Defining $X_1=G^{ij}_sG^f_{ij}$ and $X_2=(G_s^{ij}+\check H_s^{ij})^{-1}(G_f^{ij}+\check H_f^{ij})$, we can rewrite the full determinant in terms of the determinant of the solution times terms involving the fluctuations\footnote{In this and all the following intermediate expressions, there are additional terms proportional to $\vec \chi^2$ (i.e. non-derivative) that are not being written down for simplicity. I have explicitly checked that they cancel among themselves, so there is no such a term in the final fluctuation Lagrangian.} 
$$
|G_{ij}+\check H_{ij}|=|G^s_{ij}+\check H^s_{ij}| |\mathbb{I}+X_1| \sqrt{|\mathbb{I}+X_2|}
$$
To expand the last two determinants up to quadratic terms in fluctuations, we use the general formula
$$
|\mathbb{I}+X|=1+\hbox{Tr }X+\frac{1}{2}\left( \hbox{Tr }X\right)^2-\frac{1}{2}\hbox{Tr }X^2+\dots
$$
The task ahead of us is to first compute the metric and $\check H$ fluctuations, and then take the various traces. The fluctuation of the world-volume metric is
\begin{equation}
G_{ij}^f=-\frac{q\rho}{R^4}\left(\delta_i^\rho\partial_jh+\delta_j^\rho \partial_ih\right)+\frac{\rho^4}{R^4}\partial_i h \partial_j h+ \frac{R^2}{\rho^2}\partial _i \chi^m \partial_j \chi^m
\label{gfluc}
\end{equation}
We see that in the presence of non-trivial magnetic flux ($q\neq 0$) there is a term linear in fluctuations, besides the usual quadratic pieces. As for the fluctuation of the two-form $\check H^{ij}$, defined in (\ref{hhat}), we obtain
$$
\check H^{ij}_f=\check H^{ij}_s\left(1-\frac{1}{2}\hbox{Tr }X_1+\frac{1}{8}(\hbox{Tr }X_1)^2+
\frac{1}{4}\hbox{Tr }X_1^2-\frac{1}{2}G^s_{11}G^{11}_f\right)
$$
$$
-\frac{i}{3!\sqrt{-|G_s|}\sqrt{G^{11}_s}}\left(1-\frac{1}{2}\hbox{Tr }X_1\right)\epsilon^{ij1abc}f_{abc}
$$
where we recall that the indices $a,b$ run over world-volume coordinates except $\xi^1$, due to the fact that our gauge choice $a(\xi)=x^1$ for the auxiliary world-volume scalar $a(\xi)$, singles out this direction. For the same reason, in the previous expansion there is an explicit appearance of $G_{11}^s$ and $G^{11}_f$, coming from the denominator in (\ref{hhat}). To find $G^{11}_f$, we need to invert $G_{ij}^f$ in eq. (\ref{gfluc}) up to quadratic order. Explicitly,
$$
G^{11}_f=-\frac{R^6}{\rho^8R_{eff}^2}\partial_1h \partial_1h-\frac{R^{10}}{\rho^{10}}\partial_1 \chi^m \partial_1\chi^m
$$
After computing the traces and dropping total derivatives, this leads to
$$
-\sqrt{-|G+\check H|}=-\rho^3\frac{R_e^2}{R^2}\sqrt{|g_{S^3}|}
\left(1-\frac{q\rho^3}{R^4R_e^2}\partial _\rho h + 
\frac{R^2}{2\rho^2}\tilde G^{ij}\partial_i \chi^m\partial _j\chi^m+\frac{\rho^4}{2R^2R^2_{eff}}\tilde G^{ij}\partial_ih\partial_jh+\right .
$$
$$
\left.+\frac{1}{3!}\frac{R_{eff}^2}{2R^2}\tilde G^{ad}\tilde G^{be}\tilde G^{cf}f_{abc}f_{def}\right)
+\frac{1}{3!}\frac{q^2\rho^6}{R^{10}R_e^{2}}\epsilon^{01bcde}\partial_b h f_{cde}
$$
where $g_{S^3}$ is the metric on a $S^3$ of unit radius and $\tilde G_{ij}$ is the metric of $AdS_3(R_{eff}/2)\times S^3(R_{eff})$
\begin{equation}
\tilde G_{ij}d\xi^id\xi^j=\frac{\rho^4}{R^4}dx_{1,1}^2+\frac{R_{eff}^2}{\rho^2}d\rho^2+R_{eff}^2d\Omega_3^2
\label{openmet}
\end{equation}
Having completed the expansion of the square root term in the action (\ref{pstact}), we turn now to the expansion of the remaing terms. Before plugging in the expressions for the fluctuating fields, it is convenient to rewrite the second term in (\ref{pstact}) as follows,
$$
\frac{\sqrt{-|G|}}{4\partial a \cdot \partial a}\partial _i a (*H)^{ijk}H_{jkl}\partial^l a=
\frac{1}{4!}\epsilon^{1abcde}H_{1ab}H_{cde}
$$
since it is then manifest that the various contributions of the world-volume metric for this term cancel each other, and we only need to expand the 3-form H. We then obtain that the last two terms of the action (\ref{pstact}) contribute 
$$
-\frac{q\rho^6\sqrt{|g_{S^3}|}}{R^6}\partial_\rho h-\frac{1}{3!}\frac{\rho^6}{R^6}\epsilon^{01bcde}\partial_bhf_{cde}+\frac{1}{4!}\epsilon^{1abcde}f_{1ab}f_{cde}
$$
to the fluctuation Lagrangian, where again we dropped total derivative terms. We have now all the contributions to the fluctuation Lagrangian; while the computation is quite tedious, the final result is reassuringly simple. For starters, we see that the linear terms in $\partial_\rho h$ cancel each other, as they should. Putting all the pieces together, the Lagrangian for the fluctuations is
$$
{\cal L}_{fluc}=-\frac{1}{2}\sqrt{-|\tilde G|}
\left(\frac{R^4}{R_{eff}^2\rho^2} \tilde G^{ij}\partial_i \chi^m\partial _j\chi^m +\frac{\rho^4}{R_{eff}^4}\tilde G^{ij}\partial_ih\partial_jh+\frac{1}{3!}f_{ijk}f^{ijk}\right)
$$
$$
+\frac{\rho^5}{R^4R_{eff}^2}h\epsilon^{01\rho cde}f_{cde}
$$
where the last term has been rewritten using the Bianchi identity for $f_{ijk}$ and dropping total derivative terms one last time. Let's comment a couple of features of this fluctuation Lagrangian. First, the kinetic terms of the various fluctuations are controlled by the metric $\tilde G$, eq. (\ref{openmet}), which differs from the induced world-volume metric $G$, eq. (\ref{indmet}). This is completely analogous to what happens for the fluctuations of world-volume fields for D-branes in the presence of non-trivial world-volume fluxes \cite{Arean:2006vg, Arean:2007nh}, as there the kinetic terms are controlled by the open string metric, rather than the world-volume metric. While we are in a regime where there are no strings, the metric $\tilde G$, eq. (\ref{openmet}) is then the analog of the open string metric.

Second, as it already happened for similar D-brane systems considered in the literature \cite{DeWolfe:2001pq, Arean:2007nh}, the fluctuations $\chi^m$ for the scalars transverse to $S^3$ in $S^7$ are not coupled to the rest of the fluctuations, making their analysis quite straightforward; on the other hand, the fluctuation $h(\xi)$ of the position of the brane is coupled to the fluctuations of the two-form $B$ on $S^3$, and we will have to consider suitable linear combinations of these fluctuations to solve the system of equations.

\subsection{Flux independence of the spectrum}
We have derived the fluctuation Lagrangian for a family of probe embeddings, labelled by $q$, the magnetic flux on the world-volume $S^3$, which also measures the amount of bending of the probe M5-brane, see eq. (\ref{probesol}). This amount of flux $q$ enters the fluctuation Lagrangian only through the effective radius, $R_{eff}$, defined in  (\ref{radieff}). In previous work \cite{Fiol:2010wf} we have already shown that there are quantities of the dual defect theory, like the defect contribution to the integrated trace anomaly, that can be derived from this Lagrangian and depend non-trivially on $q$. In what follows, we are going to argue that the mass spectrum of small fluctuations depends in a trivial way on $q$, with the interesting consequence that the spectrum of dimensions $\Delta$ of the corresponding operators in the dual theory is independent of $q$.

We are going to present two arguments for this independence, one based in the brane construction that realizes this defect CFT, and the other in the explicit form of the fluctuation Lagrangian we have just derived. For the first argument\footnote{I would like to thank Ofer Aharony for providing this argument.} it is convenient to start with the M-theory brane configuration (before taking any near horizon limit) and bring it to type IIA. There we have a single NS 5-brane along directions 013456, and stacks of N and N+k D2 branes along 012, on the two sides of the NS 5-brane. The difference on the number of branes $k$ causes a non-trivial bending on the NS-5 brane, given by $q\neq 0$ in the previous section. The 1+1 defect field theory is now a $SU(N)\times SU(N+k)$ theory with a 1+1 ${\cal N}=(4,4)$ hypermultiplet in the bifundamental, i.e. if we write the hypermultiplet as a chiral plus antichiral field $Q,\tilde Q$, $Q$ transforms in $(N,\overline{N+k})$ and $\tilde Q$ in $(\bar N, N+k)$. The relevance of this theory is that in the IR it flows to the defect CFT we are considering. Let's look now at the operators in short multiplets. In this realization, the theory is free, and schematically chiral primaries are of the form $\tilde Q XX...Q$, with $X$ scalars from the vector multiplets. It is manifest that their dimensions $\Delta$ do not depend on $k$, and therefore don't depend on $q$. Since these operators belong to short multiplets, their dimensions $\Delta$ do not get renormalized, and the fact that they don't depend on $k$ in the UV implies that they can't depend on $k$ - and therefore $q$ - in the IR.

Our second argument comes from direct inspection of the equations of motion derived from the fluctuation Lagrangian obtained in the previous subsection. It is immediate to check that while the Lagrangian fluctuation depends on $R$ and $R_{eff}$, the equations of motion depend only on $R_{eff}$, so all the $AdS_3$ masses are of the form
 $$
 m^2(l)=\frac{f(l)}{R_{eff}^2}=\frac{f(l)}{4R_{AdS}^2}
 $$
with $l$ the principal quantum number on $S^3$. This has an obvious effect on the dual theory: since the mass and the radius of the world-volume $AdS$ enter the formula for the dimensions of the dual operators 
$$
\Delta=\frac{d\pm\sqrt{(d-2p)^2+4m^2R_{AdS}^2}}{2}
$$
only through the combination $m^2R_{eff}^2$, these dimensions don't depend on $q$.

This independence on the magnetic flux of the dimensions of operators in short multiplets is not unique to the defect theory under discussion. We will now argue that it also takes place in defect CFTs coming from D-brane setups. For the D3-D5 system introduced in \cite{Karch:2000gx} and extensively discussed in \cite{DeWolfe:2001pq}, introducing a non-zero magnetic flux on the world-volume of the probe D5, corresponds to having some $k\neq 0$ D3s ending on the D5 brane, in the brane configuration. The question again turns into the (in-)dependence of the dimensions $\Delta$ of the operators in short multiplets on this new integer $k$. Again we can present two arguments, very similar to the ones above. The first uses the S-dual brane configuration, where the D5 brane turns into an NS5-brane, and the defect theory is a 2+1 quiver theory with $SU(N)\times SU(N+k)$ gauge group and a hypermultiplet in the bifundamental. Again in the free limit (which we can take, since we know that the dimensions don't get renormalized) it is clear that the dimensions of these operators don't depend on $k$, so they don't depend on $q$ for any value of the coupling.  The second way to reach the same conclusion is again by direct inspection of the equations of motion for the fluctuations. The corresponding fluctuation Lagrangian for $q\neq 0$ can be read from \cite{Arean:2007nh} to be
$$
{\cal L}_{fluc}=-\rho^2 \sqrt{g_{S^2}}\left[\frac{\rho^2}{2R^2}\tilde G^{ab}\partial_a h\partial _b h+\frac{R^2_{eff}}{2\rho^2}\tilde G^{ab}\partial_a\chi^m\partial_b\chi ^m+\frac{R^2_{eff}}{4R^2}f_{ab}^2\right]-\frac{4\rho^3}{R^2R^2_{eff}}hf_{\theta \phi}
$$
 where $\tilde G$ is the open string metric $AdS_4(R_{eff})\times S^2(R_{eff})$ world-volume metric
 $$
 G_{ij}d\xi^id\xi^j=\frac{\rho^2}{R^2}dx_{1,2}^2+\frac{R_{eff}^2}{\rho^2}d\rho^2+R_{eff}^2d\Omega_2
 $$
 and $R_{eff}$ is given by
 $$
 R_{eff}^2\equiv R^2\left(1+\frac{q^2}{R^4}\right)
 $$
 It is again immediate to check that the equations of motion depend only on $R_{eff}$, so again all $AdS_4$ masses depend only on this scale, and the dimensions of the dual operators are independent of $q$.
 
\subsection{The fluctuation spectrum}
Having obtained the fluctuation Lagrangian, our next task is to solve the resulting equations of motion.
As usual, this is achieved by performing a mode decomposition over the world-volume $S^3$, in terms of the relevant $S^3$ spherical harmonics. This allows us to obtain the $AdS_3$ masses of all the modes.

Since we argued in the previous subsection that the dimensions of operators in short multiplets are independent of $q$, we will compute them setting $q=0$ in the fluctuation Lagrangian,
$$
{\cal L}=-\frac{1}{2}\rho^3 \sqrt{|g_{S^3}|} \left(\frac{\rho^4}{R^4}G^{ij}\partial _i h\partial _j h +\frac{R^2}{\rho^2}G^{ij}\partial_i \chi ^m \partial _j \chi ^m +\frac{1}{3!}f_{ijk}f^{ijk}\right)
+\frac{6\rho^5}{R^6}h f_{\theta_1\theta_2\theta_3}
$$

\underline{Fluctuations in $S^7$}.
The fluctuations $\chi^m$ of the four scalars transverse to $S^3$ in $S^7$ decouple from the rest, and the corresponding equation of motion is
$$
\partial_i\left(\rho \sqrt{|g_{S^3}|} G^{ij} \partial_j \chi ^m\right)=0
$$
The geometric content of this equation is manifest by switching to angular variables. For small fluctuations this amounts to the field redefinition $\chi^m=\rho \Psi^m$, which allows to rewrite this equation as
$$
\left( \Box _{AdS_3}+\frac{1}{R^2}\Box_{S^3}+\frac{3}{R^2}\right)\Psi^m=0
$$
Separating variables by writing $\Psi(\xi)=e^{ikx}A(\rho)Y_l(S^3)$, and using the spectrum of scalar harmonics $Y_l(S^3)$ on $S^3$, we deduce that the $AdS_3$ masses for each of the four $\chi^m$ are
$$
m^2(l)=\frac{-3+l(l+2)}{R^2}=\frac{-3+l(l+2)}{4R_{AdS}^2}
$$
The $l=0$ mode is tachyonic, $m^2(l=0)=-3/R^2$, but it does not violate the $AdS_3$ BF bound
$m^2_{BF}=-1/R_{AdS_3}^2$, since $R_{AdS^3}=R/2$ and therefore $m^2(l=0)=-3/4R_{AdS_3}^2$.
This tachyon is a 'slipping' mode that reflects the possibility for the brane of wrapping a non-maximal $S^3$ in $S^7$. The modes with $l=1$ are massless; this masslessness reflects the freedom to rotate our choice of equatorial $S^3$ inside $S^7$.

\underline{Coupled fluctuations}.
The equation of motion for the $x^2$ fluctuation $h(\xi)$ is
$$
-\partial_i\left(\rho^7\sqrt{g_{S^3}}\tilde G^{ij}\partial_j h\right)=\frac{6\rho^5}{R^2}f_{\theta_1\theta_2\theta_3}
$$
In order to realize the geometric content of this equation, it is convenient to rewrite it in terms of $\psi=\rho^2 h$, 
$$
\left(\Box_{AdS}+\frac{1}{R^2}\Box_{S}-\frac{12}{R^2}\right)\psi=-\frac{6}{\sqrt{g_{S^3}}R^2}f_{\theta_1\theta_2\theta_3}
$$
We turn now to the equations of motion for the two-form $b_{ij}(\xi)$, 
$$
\partial_i \left(\rho^3\sqrt{g_{S^3}}f^{ijk}\right)=\frac{6\rho^5}{R^6}\epsilon^{01\rho ijk}{\partial _i h}
$$
We observe that the $h$ fluctuation and the fluctuations of the $B$ field over $S^3$ are coupled. To decouple them we consider fluctuations over $S^3$ with the following ansatz 
$$
b_{\alpha \beta}(\xi)=\sqrt{g_{S^3}}\epsilon_{\alpha \beta \gamma} g^{\gamma \delta}\partial _\delta \phi(\xi) 
$$
The system of equations now reduces to
$$
\left(\Box_{AdS}+\frac{1}{R^2}\Box_{S^3}-\frac{12}{R^2}\right)\psi=-\frac{6}{R^2}\Box \phi
$$
$$
\partial_\alpha \left(\Box_{AdS}+\frac{1}{R^2}\Box_{S^3}\right)\phi=\frac{6}{R^2}\partial_\alpha \psi
$$
Since the fields $\Psi(\xi)$ and $\phi(\xi)$ are scalars over $S^3$ we decompose them in scalar spherical harmonics. It proves convenient to treat the constant mode ($l=0$) on $S^3$ separately, since it is immediate that in this case the second equation above is trivially satisfied. For $\l>0$ the system of equations can be easily diagonalized, by taking the linear combinations $Z^+=\psi+l\phi$ and
$Z^-=\psi-(l+2)\phi$. The resulting spectra of masses for the decoupled modes are
$$
m^2(l)_+=\frac{(l+6)(l+2)}{R^2}
$$
$$
m^2(l)_-=\frac{l(l-4)}{R^2}
$$
There are two towers of modes, related by $m_+^2(l)=m_-^2(l+6)$. In the first tower all the modes are massive. On the other hand, for the second tower of modes, when its masses are written in terms of $R_{AdS_3}=R/2$ we have
$$
m^2(l) R_{AdS_3}^2=\frac{l(l-4)}{4}=-\frac{3}{4},-1,-\frac{3}{4},0,\dots
$$
so the most tachyonic mode corresponds to $l=2$, which actually saturates the $AdS_3$ BF bound. 

Coming back to the $l=0$ case, the second equation becomes trivial, and the first one is immediately solved, yielding $m^2(l=0)=12/R^2$, which is the value one obtains for the $m^2_+(l)$ tower setting $l=0$. So as it is common in these KK reductions, one of the two branches actually extends to $l=0$.

\underline{One-forms on $AdS_3$}. We can take the remaining two modes of the 2-form to have one index in $AdS_3$ and the other in $S^3$, so we write them as a sum of one-forms on each,
$$
b_{\mu\alpha}^\pm=\sum_l b_\mu^l \epsilon_{\alpha}^{\beta \gamma}
\nabla_{[ \beta}Y_{\gamma ]}^{l\pm}
$$
where $Y^{\pm}_\alpha$ are two of the possible three vector harmonics on $S^3$.  For given $l$, the various $b_\mu^l$ are not independent, but rather related by the various equations of motion. This is most transparent if we switch to a description in terms of forms as in \cite{DeWolfe:2001nz}, so the resulting condition reads $d^\dagger_{AdS_3} b=0$. Using this constraint, the remaining equation can be written as
$$
\left(\Delta -\frac{(l+1)^2}{R^2}\right)b^l=0
$$
where $(l+1)^2=(l+1)(l+3-2)$ are the eigenvalues of the vector spherical harmonics on $S^3$ with respect to the Hodge-deRham Laplacian. The resulting masses are then
$$
m^2(l)=\frac{(l+1)(l+3-2)}{R^2}=\frac{(l+1)^2}{4R^2_{AdS}}
$$

\subsection{Dimensions of dual operators}
Before adding the M5 brane, the 2+1 CFT is the IR fixed point of 2+1 ${\cal N}=8$ SU(N) SYM. The work of ABJM \cite{Aharony:2008ug} provided an explicit Lagrangian for this theory (their $k=1$ case), which is expected to be dual to M-theory on $AdS_4\times S^7$. Once we add the M5-brane, we have a defect 1+1 non-chiral CFT with ${\cal N}=(4,4)$ SUSY, with a 1+1 ${\cal N}=(4,4)$ hypermultiplet living on the defect. Using the results of the previous subsection, and the standard AdS/CFT relation for the dimensions of operators dual to $p$-forms
\begin{equation}
\Delta=\frac{d\pm\sqrt{(d-2p)^2+4m^2R_{AdS}^2}}{2}
\label{reladim}
\end{equation}
we can compute the dimensions of the operators dual to these modes. Non-trivially, they turn out to be all half-integer, which is a clear sign that they belong to short multiplets. In fact, this feature can be anticipated by again going to the type IIA setup discussed in subsection 3.1. In the D2-NS5 configuration the chiral primaries are operators of the form $\tilde QXX..Q$ and since $\Delta(Q)=\Delta(\tilde Q)=0$ and $\Delta(X)=1/2$, we see that in the UV all dimensions in short multiplets are half-integer; since they don't get renormalized, the same is true in the IR.

The relation (\ref{reladim}) in principle allows for two dimensions for a given value of $m^2$. We must also require that the dimensions satisfy the corresponding unitarity bounds. Although we are considering a 1+1 defect theory, as in \cite{Constable:2002xt}, we found no signs of a Virasoro algebra, so we impose the unitarity bounds of the two dimensional global conformal symmetry $SL(2,\mathbb{C})$.

The transverse fluctuations are four scalars on $AdS_3$. The dimensions of the dual operators are
$$
\Delta_+=\frac{l+3}{2}\hspace{1cm}\Delta_-=\frac{1-l}{2}
$$
The relevant unitarity bound $\Delta_-\geq 0$ implies that this is only possible for $l=0,1$.

The modes in the coupled system are scalars. For the first tower we have
$$
\Delta_{+}^{(+)}=\frac{l+6}{2} \hspace{1cm}\Delta_{-}^{(+)}=-\frac{l+2}{2}
$$
so $\Delta_-$ is never allowed. For the second tower we have
$$
\Delta_{+}^{(-)}=\frac{l}{2}\hspace{1cm}\Delta_{-}^{(-)}=\frac{4-l}{2}
$$
and $\Delta_-$ is allowed for $l=1,2,3,4$.

Finally, the remaining $b$ modes are 1-forms in $AdS_3$, so using the mass/dimension relation (\ref{reladim}) for 1-forms, we get
$$
\Delta_+=\frac{l+3}{2}\hspace{1cm}\Delta_-=\frac{1-l}{2} 
$$
and $\Delta_-$ is never allowed.

\section{Meson spectrum}
So far, we have considered a defect conformal field theory. We want to study the effect of breaking conformal invariance, by giving a mass to the hypermultiplet. This produces a discrete set of mesonic states, restricted to the 1+1 defect. The effect of giving a mass to the hypermultiplet of this system has already been considered in \cite {Arean:2006pk, Myers:2006qr,  Arean:2007nh}, but only for the fluctuations $\chi^m$ in $S^7$. By supersymmetry, this is actually enough to determine the masses of all the mesons, since mesons in the same multiplet must have the same mass. We will extend this analysis by providing the equations of motion for the mesons coming from the remaining fluctutations. This could be useful if one is interested in a more detailed analysis of these mesons, beyond just their mass spectrum, for instance if one is interested in computing their form factors. 

It was argued in \cite {Arean:2007nh} that the discrete meson spectrum is a distinctive feature of the system with $q=0$, and a non-zero $q$ renders again the spectrum continuous and gapless. For this reason, we will consider only the case of zero magnetic flux $q=0$. In the M-brane setup, a non-zero mass for the hypermultiplet corresponds to placing the probe M5-brane at a distance $|\vec z|=L$ from the stack of M2-branes, i.e. the new ansatz is 
$$
x^2=x^2(\rho),\hspace{1cm}|\vec z|=L, \hspace{1cm} F=0
$$
It was shown in \cite{Arean:2007nh} that $x^2=x^2(\infty)$ still solves the equations of motion. The induced world-volume metric is now
$$
ds^2_{M5}=\frac{(\rho^2+L^2)^2}{R^4}dx^2_{1,1}+\frac{R^2}{\rho^2+L^2}\left(d\rho^2+\rho^2 d\Omega_3^2\right)
$$
The world-volume metric is no longer $AdS_3\times S^3$, it only approaches this form asymptotically as $\rho\rightarrow \infty$. As a result of $L\neq 0$, besides breaking conformal invariance of the defect theory, the internal symmetry group gets reduced from $SO(4)\times SO(4)$ to $SO(4)\times SO(3)$.

The derivation of the fluctuation Lagrangian can be carried out following the same steps as in subsection 3.1, and it is actually much simpler, due to the absence of magnetic flux. We omit the details and present only the final result. Since we are considering a situation without magnetic flux, the kinetic terms of the fluctuation Lagrangian are controlled by the induced world-volume metric (the M-theory analogue of the open string metric now coincides with the world-volume metric),
$$
{\cal L}_{fluc}=-\frac{1}{2}\sqrt{-|G|}\left(G^{ij}\partial_ih\partial_jh\frac{(\rho^2+L^2)^2}{R^4}+G^{ij}\partial _i\chi^m \partial_j\chi^m\frac{R^2}{\rho^2+L^2}+\frac{1}{3!}f_{ijk}f^{ijk}\right)
$$
$$
+\frac{\rho (\rho^2+L^2)^2}{R^6}h\epsilon^{01\rho cde}f_{cde}
$$
From this Lagrangian we can derive the equations of motion. Separation of variables always turns them into a one-dimensional problem in the radial direction $\rho$. Imposing regularity and normalizability  of the wave function we derive a discrete set of modes with masses $M(n,l)$, where $n$ is the radial number, and $l$ is the principal quantum number on $S^3$.

\underline{Mesons from $S^7$ fields}. Their equation of motion is
$$
\frac{1}{\rho}\partial_\rho(\rho^3\partial_\rho \chi^m)+\frac{R^6\rho^2}{(\rho^2+L^2)^3}\partial_\mu \partial^\mu \chi^m+\Box_{S^3}\chi^m=0
$$
Separating variables by writing $\chi +e^{ikx}f(\rho)Y^l(S^3)$ and denoting $M^2_s=-k^2$ we have
$$
\frac{1}{\rho}\partial_\rho(\rho^3\partial_\rho f)+\frac{R^6M^2_s\rho^2}{(\rho^2+L^2)^3}f-l(l+2)f=0
$$
This equation has already been derived in the literature \cite{Arean:2006pk, Myers:2006qr}. For $l=0$, it is possible to solve it analitycally in terms of Bessel functions, and demanding that the wavefunctions are regular at the origin and normalizable, we arrive at the conclusion that the masses $M_s(n,0)$ are given by the zeros of $J_1$,
$$
J_1\left(\frac{R^3M_s(n,0)}{L^2}\right)=0
$$
For $l>0$, I am not aware of analytic solutions of this equation. One can then resort to numeric analysis, but it is possible to estimate the discrete masses for large radial number $n$, using the WKB approximation (see  \cite{Russo:1998by} for a detailed exposition of the WKB approximation for these type of problems) ; in this case one gets
$$
M^{WKB}_s(n,l)=\frac{\pi L^2}{R^3}\sqrt{(n+1)(n+\frac{3}{2}(l+1))}
$$
\underline{Mesons from the coupled sector}.
As it happened in the conformal case, fluctuations of $x^2$ couple to fluctuations of $B$ on $S^3$.
Introducing the same ansatz we used for the conformal case we arrive at the system of equations
$$
\frac{R^4\rho^2}{(\rho^2+L^2)^3}(-\partial_0^2+\partial_1^2)h+\frac{1}{R^2}\frac{1}{\rho(\rho^2+L^2)^3}
\partial_\rho\left[\rho^3(\rho^2+L^2)^3\partial_\rho h\right]+\frac{1}{R^2}\Box h=
-\frac{6}{R^2}\frac{f_{\theta_1\theta_2\theta_3}}{(\rho^2+L^2)\sqrt{g_{S^3}}}
$$
$$
\left(\frac{R^4\rho^2}{(\rho^2+L^2)^3}(-\partial_0^2+\partial_1^2)+\frac{1}{R^2}\frac{\rho^3}{(\rho^2+L^2)^3}\partial_\rho\left[\frac{(\rho^2+L^2)^3}{\rho}\partial_\rho \right]+\frac{1}{R^2}\Box\right)\partial_\alpha \phi=
\frac{6}{R^2}\frac{\rho^4}{(\rho^2+L^2)}\partial_\alpha h
$$
 Remarkably, the same linear combination that decouples this system in the conformal ($L=0$) case, continues to decouple it for $L\neq 0$. This feature is also present in defect theories coming from D-branes \cite{Arean:2006pk}. Introducing the same linear combinations $Z_{\pm}$ as in the conformal case, and separating variables as $Z_\pm=e^{ikx}f(\rho)_\pm Y(S^3)$, we are left with a single equation for $f_\pm(\rho)$
$$
\partial_\rho\left(\frac{(\rho^2+L^2)^3}{\rho}\partial_\rho f_\pm\right)+\frac{R^6M^2}{\rho}f_\pm-\frac{(\rho^2+L^2)^3}{\rho^3}l(l+2)f_\pm-\frac{(\rho^2+L^2)^2}{\rho}6(-1\pm(l+1))f_\pm=0
$$
We can actually argue that is equivalent to the equation for mesons coming from fluctuations in $S^7$. To do so, we follow a procedure outlined in \cite{Myers:2006qr} in their study of the meson spectrum of the $Dp-D(p+4)$ system. For concreteness consider the case $l=L-1$. We first introduce a function $F(\rho)$ by $\chi=\rho^{L-1}F(\rho)$ and rewrite the equation of motion of $\chi$ in terms of $F$. If we derive the resulting equation with respect to $\rho$, we obtain an equation that only depends on $F$ through its derivatives. Defining $G(\rho)=\rho^{L+1}\partial _\rho F(\rho)$ we arrive at a differential equation for $G(\rho)$ which is actually the one above. By a similar argument for the plus sign, we deduce that 
$$
M_\pm(n,l)=M_s(n,l\pm 1)
$$
\underline{Mesons from the 2-form}. As we already did in the conformal case, we expand the 2-form fluctuation with one index in $S^3$ as
$$
b_{\mu \alpha}^{\pm}=\sum _l b_\mu \epsilon_\alpha^{\beta \gamma} \nabla_{[\alpha} Y_{\beta ]}^{l\pm}
$$
When $L\neq 0$, is no longer true that the world-volume metric is of product form, since now the radius of the world-volume $S^3$ depends on $\rho$. This barely complicates the analysis, which proceeds along the lines of the $L=0$ case. The fields $b_\mu$ are not independent, but rather satisfy $d^\dagger \left (\frac{(L^2+\rho^2)^{1/2}}{\rho} b\right)=0$,  which reduces to the condition $d^\dagger b=0$ of the conformal limit. Using this relation in the remaining equations of motions, we arrive at a equation for $b_\rho$ alone, namely
$$
\frac{\rho^2}{(\rho^2+L^2)^3}\partial_\rho
\left[\rho\partial_\rho\left(\frac{(\rho^2+L^2)^3}{\rho}\right)b_\rho \right]
+\frac{R^6M^2_1\rho^2}{(\rho^2+L^2)^3}b_\rho-(l+1)^2b_\rho=0
$$
This equation can be shown to be identical to the one determining the spectrum of mesons from transverse fields by the change of variables $\chi=(\rho^2+L^2)^3 b_\rho/\rho^2$, so these modes give the same masses,
$$
M_1(n,l)=M_s(n,l)
$$

To write this meson spectrum in field theory parameters, we need to relate the mass of the hypermultiplet to the separation $L$ between the $M5$-brane and the stack of $M2$-branes. This was already done in \cite{Myers:2006qr}, who argued that $m\simeq L^2/l_P^3$, and therefore using the relation
$$
\frac{R}{l_P}=\left( 32\pi^2N_c\right)^{1/6}
$$
concluded that the masses of the mesons behave like $M\simeq m/N_c^{1/2}$, i.e. these mesons are deeply bound.

\section{Acknowledgements} 
I would like to thank Ofer Aharony for insightful discussions and Alfonso Ramallo for useful correspondence regarding \cite{Arean:2007nh}. I would like to thank the Department of Particle Physics at the Weizmann Institute of Science and the Theory group at NIKHEF (Amsterdam) for hospitality at various stages of this project. This research is supported by a Ram\'{o}n y Cajal fellowship, and also by MEC FPA2007-66665C02-02, CPAN CSD2007-00042, within the Consolider-Ingenio2010 program, and AGAUR 2009SGR00168.

\end{document}